\documentstyle[aps,prl,epsf]{revtex}
\draft
\tighten

\preprint{Version 2.0}

\begin{document}

\title{Decay rates and the hyperfine structure of the
bound $\mu^+\mu^-$ system}

\author{U. D. Jentschura, G. Soff\footnote{Electronic address:
ulrich@theory.phy.tu-dresden.de}} 
\address{Institut f\"ur Theoretische Physik, Technische Universit\"{a}t 
Dresden, 01062 Dresden, Germany
\footnote{temporary address}}

\author{V. G. Ivanov}
\address{Pulkovo Observatory, 196140 St. Petersburg, Russia}

\author{S. G. Karshenboim\footnote{Electronic address:
sgk@onti.vniim.spb.su}}
\address{D. I.  Mendeleev Institute for Metrology,
198005 St.~Petersburg, Russia\footnote{permanent address},\\
Max--Planck--Institut f\"{u}r Physik komplexer
Systeme, 01069 Dresden, 
Germany\footnote{temporary address}}

\maketitle

\begin{abstract}
Results are obtained
for the decay rate of ortho and para states 
and for the hyperfine structure
of the dimuonic system $\mu^{+}\mu^{-}$ (dimuonium). 
We calculate next--to--leading order
radiative corrections.
The decay rate is strongly influenced by 
the electronic vacuum polarization in
the far time--like asymptotic region 
and thus allows for a test of QED in a 
previously unexplored kinematic regime.  
\end{abstract}


\noindent
The production of exotic atoms and their properties 
have been discussed in a number of recent publications 
(see e.g. \cite{pith1,pith2,pith3,pipi2}).  
Pionium, the bound system of a positive and a negative pion,
has been observed 
in a beam target experiment with 
proton projectiles at the Serpukhov accelerator \cite{pipi2}.
Atomic bound states were also observed in decay modes of certain 
particles, e.g. positronium can be formed in the $\pi^0 \to
{\sl positronium}+\gamma$ decay of the neutral pion \cite{posgamdecay},
and the $\pi\,\mu$ atom can be produced in the reaction
$K_L^0 \to \left(\pi \, \mu\, {\sl atom}\right) + \nu$
\cite{pimudecay}.

The bound system consisting of two muons (dimuonium)
can be produced in
heavy nuclei inelastic scattering at high energies
and in particle decays.
The decay of the neutral $\eta^0$ meson into dimuonium
has been investigated theoretically
by L. Nemenov \cite{nemenov}
($\eta^0 \to {\sl dimuonium} + \gamma$).
The formation of dimuonium in pion--proton collisions
($\pi^{-} + p \to {\sl dimuonium} + n$) and
by photons on nuclei ($\gamma + Z \to {\sl dimuonium} +
Z$) has been discussed by S. Bilenkii {\em et al.} 
\cite{bilenkii}.
For the direct production of dimuonium in muon--antimuon
collisions, experimental difficulties
associated with slow muon beams would have to be overcome.
Another possible pathway for the production of the system,
which we do not discuss in any further detail here, is the
$e^+e^-$-annihilation
(near or above the $\mu$ threshold).
Dimuonium, once produced, undergoes
atomic decay (into energetically lower atomic states) and
annihilation decay (into electrons and photons).
Because the annihilation products are
hard photons and relativistic
electron--positron pairs, the decays could be
investigated experimentally by established methods
of particle physics.

The lifetimes of low-lying $S$ states in the dimuonic system
lie in the $10^{-12}\,{\rm s}$ 
range. The decay products are hard photon pairs
and relativistic electron-positron pairs, which can be easily detected.
Because decay rate measurements can usually 
be performed with only a small number of
events available, it is natural to begin the study of exotic
systems with a detailed analysis of the decay channels and rates.

In this Letter we obtain results for 
the decay rates of the $n^3S_1$ ortho and the
$n^1S_0$ para states of the dimuonic system for $n=1$ and $n=2$.
We evaluate radiative corrections in 
next--to--leading order. We analyze the hyperfine structure 
of the system and obtain results for $n=1$ and $n=2$ states.

The name ``dimuonium'' was proposed
by J. Malenfant for the bound system of two muons
in his pioneering investigations on the system 
\cite{malenfant1,malenfant2}.
In analogy to positronium we speak of ortho-- and
paradimuonium. We observe that a precise
measurement of the orthodimuonium
decay rate would allow for sensitive tests of QED. 
We emphasize
that both the hyperfine structure and the decay rate of dimuonium
are influenced by the electronic vacuum polarization 
in the far time--like
asymptotic region (due to virtual photon annihilation processes). 
This regime does not yield any contribution in any other muonic 
or electronic bound system (see e.g. \cite{mohr}, \cite{jungmann}). 
The electronic vacuum polarization
contributes a relative correction of $1.6\,\%$ to the decay rate 
of orthodimuonium, and thus would allow a test of QED in a 
previously unexplored kinematic regime.

The hadronic contribution to the vacuum polarization 
(for time--like $q^2$ below the pion threshold) 
can be found from a precise measurement of the
orthodimuonium decay rate or the hyperfine structure. The 
correction enters
at the level of one permille.  

\begin{figure}[htb]
\centerline{\mbox{\epsfysize=10cm\epsffile{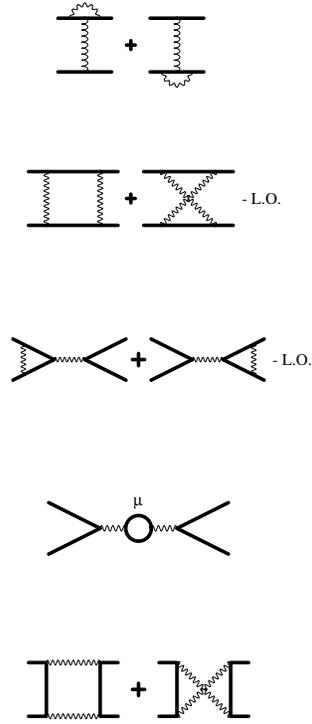}}}
\caption{\label{diagPs} Known corrections to the hyperfine splitting of
$S$ states of a dileptonic system. 
Springy lines denote transverse photons.}
\end{figure}

\begin{figure}[htb]
\centerline{\mbox{\epsfysize=10cm\epsffile{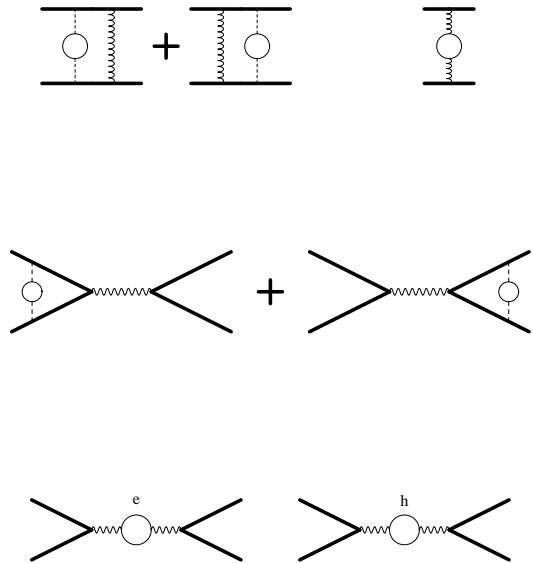}}}
\caption{\label{diag2mu} Additional corrections to the hyperfine
splitting specific to dimuonium. Bold fermionic lines
denote muons, thin lines denote electrons. Dashed lines denote Coulomb 
photons. }
\end{figure}

\begin{figure}[htb]
\centerline{\mbox{\epsfysize=12cm\epsffile{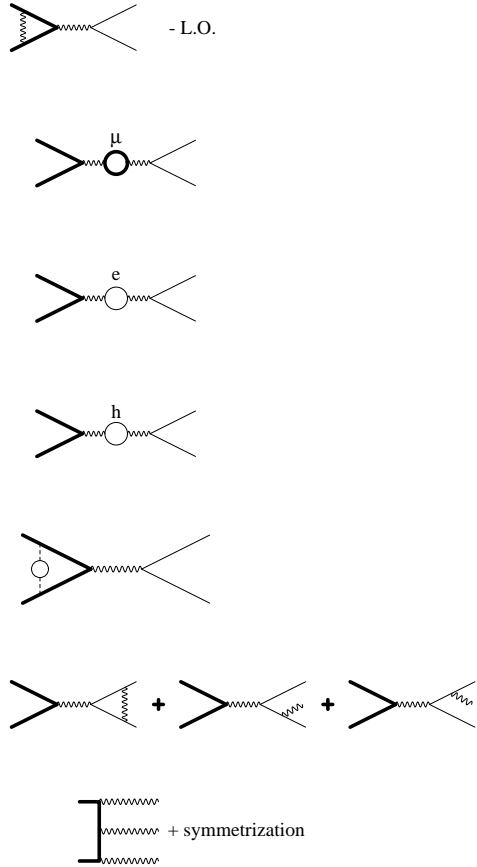}}}
\caption{\label{diagOM} Feynman diagrams for next--to--leading order
corrections to the decay rate of orthodimuonium. The subtraction of lower order (L.O.) terms is necessary for
some contributions in order to prevent double counting.}
\end{figure}

\begin{figure}[htb]
\centerline{\mbox{\epsfysize=6cm\epsffile{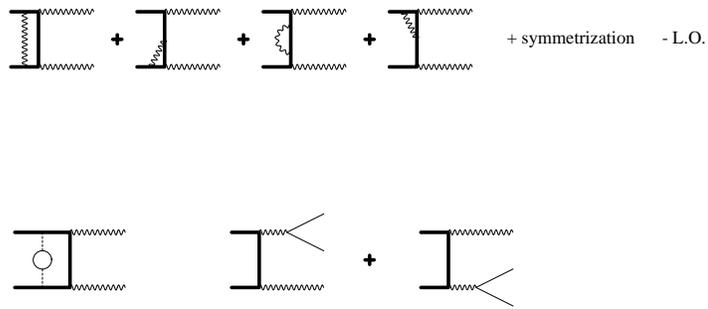}}}
\caption{\label{diagPM} Feynman diagrams for next--to--leading order
corrections to the decay rate of paradimuonium.}
\end{figure}

We study here radiative corrections to the orthodimuonium
and paradimuonium decay (Fig. \ref{diagOM} and \ref{diagPM}, 
respectively).  We 
also investigate the hyperfine splitting (hfs) 
of the ground state 
(Fig. \ref{diagPs} and Fig. \ref{diag2mu}).
Some of the contributions to the hyperfine splitting are
closely related to corrections for the orthodimuonium 
decay, and certain corrections can be directly found in analogy
with the positronium hyperfine splitting (Fig. \ref{diagPs}),
which has been considered e.g. 
in \cite{IZ,KK}. The leading term for the hyperfine 
splitting is the Fermi energy $E_F = (7/12)\,\alpha^4\,m$,
where $m=m_e$ for positronium and $m = m_\mu$ for dimuonium
($\hbar = c = 1$).
We list here the next--to--leading order contributions 
to hfs in the order in which they are depicted in Fig. \ref{diagPs}.
They are due to the vertex correction (fermion line), 
recoil corrections, vertex corrections in the annihilation
part, vacuum polarization due to a loop of the constituent
fermion ($e^+e^-$ for positronium, $\mu^+\mu^-$ for dimuonium),
and virtual two--photon--annihilation. These contributions,
respectively, are given by (see \cite{IZ,KK,BS})  
\begin{eqnarray}  \label{ps5}
E^{(5)}_{\rm Ps}(nS) &=& 
\frac{\alpha}{\pi} \left[ \frac{4}{7} - \frac{6}{7} +
\frac{12}{7} - \frac{8}{21} + \left( - 
\frac{6}{7}\,\ln 2 - \frac{6}{7} + \frac{3}{7}\, \pi\,i \right)\right] 
\cdot \frac{E_F}{n^3} \nonumber\\
&=& \frac{\alpha}{\pi} \left[ -
\frac{6}{7} \ln 2 - \frac{32}{21} + \frac{3}{7}\,\pi\,i \right] 
\cdot \frac{E_F}{n^3} \,.
\end{eqnarray}
For dimuonium, the small length scale of the system enhances the
effects of the vacuum polarization corrections. The {\em electronic}
vacuum polarization generates corrections to the wave function
which enter at the order of $\alpha/\pi$ times the leading term 
as well as a large
contribution due to an electron--positron vacuum polarization 
loop in the 
virtual annihilation diagram. The additional
corrections specific to dimuonium, which are not present in
positronium, are depicted in Fig.
\ref{diag2mu}. The contributions are given by the (electronic) 
vacuum polarization corrections to the wave function
in the transverse photon exchange diagram, by a 
vacuum polarization insertion in the transverse photon line,
by vacuum polarization corrections to the wave function in the 
annihilation diagram, and by electronic and hadronic vacuum
polarization insertions in the virtual photon line 
(annihilation diagram). Respectively, we obtain the following results
for the additional contributions specific to dimuonium,
\begin{equation}  \label{dimu5}
E^{(5)}_{\mu^+\mu^-}(1S) = 
\frac{\alpha}{\pi} \, \left[ 0.605 + 0.345 + 0.454 + 
\left(\frac{2}{7} \, \ln \frac{2\,m_\mu}{m_e} - \frac{5}{21} -
\frac{1}{7} \, \pi \,i\right)
-0.080(9) \right] \cdot E_F\,. 
\end{equation}
It should be noted that the
first three contributions in Eq. (\ref{dimu5}) are state
dependent, i.e. they do not exhibit the familiar $1/n^3$ scaling.
Because atomic momenta are of the order $\alpha\,m_\mu$, it is
sufficient to carry out the calculations with non-relativistic
wave functions (cf. \cite{karshenboim1}). The numerical methods
employed in \cite{karshenboim1} for the treatment of the non-relativistic
Green's function are also used for the calculations presented in
this Letter. The results have been checked by independent evaluations
using the computer algebra system {\sc Mathematica} \cite{mathematica},
and will be described further in \cite{pre}.

In the discussion of the hadronic vacuum polarization, we follow an
approach outlined in \cite{sapirstein1}. We thereby estimate the 
uncertainty due
to our model as $11\,\%$ of the total hadronic vacuum polarization 
contribution. This is sufficient because the hadronic contribution is
small compared to other corrections, and because
higher order QED radiative corrections
are expected to contribute in the same order as
the uncertainty of our model of hadronic vacuum polarization.
The main contribution to the hadronic vacuum polarization originates
from the $\rho$ meson pole in the pion form factor.
 
The real part of the 
sum of all the next--to--leading order corrections for dimuonium is 
\begin{equation} \label{hfs1}
E^{(5)}_{\rm hfs}(1S) = E^{(5)}_{\rm Ps} +
E^{(5)}_{\mu^+\mu^-} =
\frac{\alpha}{\pi} \, 0.689(9) \cdot E_F\,.
\end{equation}
For the $2S$ state, the contributions sum up to
$E^{(5)}_{\rm hfs}(2S) = (\alpha/\pi) \, 0.556 \cdot E_F/8$.
We obtain the following theoretical values for dimuonium
hyperfine splitting,
$E_{\rm hfs}(1S) =  4.23283(35)\cdot 10^{7}\,{\rm MHz}$
and $E_{\rm hfs}(2S) = 5.28941(34) \cdot 10^{6}\,{\rm MHz}$,
where we estimate higher order quantum electrodynamic corrections
to enter at the level of $5\,\%$ of the result obtained for
next--to--leading order corrections. The hyperfine structure of dimuonium
could be measured in an infrared laser field which mixes 
the two states and thus
modifies the decay rate of the statistical sample.

The main decay channel for orthodimuonium is given by one--photon
annihilation into an electron--positron pair. The
first--order result can be extracted 
from the imaginary part of the electronic vacuum polarization
contribution (annihilation diagram) to the hyperfine splitting
(Eq. (\ref{dimu5})). The result is
$\Gamma^{(0)}(n{^3}S_1) = \alpha^5 \, m_\mu/(6\,n^3)$.
Corrections to this result (see Fig. \ref{diagOM}) 
originate from similar diagrams
as those for the hyperfine splitting. Some
of the calculations are therefore closely related to those
for the hyperfine splitting.
The next--to--leading order corrections are due to
radiative corrections to $\mu$ lines, 
due to the muonic, electronic and hadronic vacuum
polarization, due to the wave function correction 
caused by electronic vacuum polarization,
due to radiative corrections in the electron--positron--line and
bremsstrahlung corrections, and three-photon decay.
We obtain for the respective contributions,
\begin{eqnarray} \label{ortho2} 
\Delta \Gamma(1{^3}S_1) &=& \frac{\alpha}{\pi} \,
\left[ -4 - \frac{16}{9} + 
\left(\frac{4}{3}\, \ln \frac{2\,m_\mu}{m_e} - \frac{10}{9} \right)
- 0.37(4) + 1.06 + \frac{3}{4} + \frac{4}{3}\,\left(\pi^2 - 9\right) 
\right] \cdot \Gamma^{(0)}(1{^3}S_1) \nonumber\\
&=& \frac{\alpha}{\pi} \, 3.74(4) \cdot \Gamma^{(0)}(1{^3}S_1)\,.
\end{eqnarray}
For the $2S$ state, we obtain the result
$\Delta \Gamma(2{^3}S_1) = (\alpha/\pi) \, 3.60(4) \cdot 
\Gamma^{(0)}(2{^3}S_1)$.

Corrections to the orthodimuonium decay rate have been investigated
previously \cite{malenfant1,malenfant2}. In \cite{malenfant2},
only the decay of orthodimuonium into a single electron-positron pair 
and into an electron-positron pair and a photon is taken into consideration.
Three--photon decay, however, constitutes an important correction
to the overall decay rate (last term in Eq. (\ref{ortho2})). 

There is also
a numerical discrepancy between the result reported 
in \cite{malenfant2} and our result for 
the contribution of the electronic vacuum polarization
correction to the wave function 
(fifth diagram in Fig. \ref{diagOM}, fifth term in Eq. (\ref{ortho2})).
We presume this 
disagreement can be traced to the fact that J. Malenfant has 
calculated this
correction with free Green functions, whereas the evaluation
in this work is done using bound Green functions.
We hold the view that it is indispensable to use
bound Green functions for the respective
correction, because the atomic
momentum in dimuonium is of the order
of $\alpha \, m_\mu/2 \approx 0.75 \, m_e$.
This momentum is close to
the mass of the loop particles (electrons and positrons) 
of electronic vacuum polarization.
These particles determine the radius of the Uehling potential.
Therefore, the decisive region of the momentum integration 
for this
correction is about $m_e$, 
where the bound Coulomb Green function cannot be approximated
by the free Green function (because the effect of
the binding Coulomb potential, in momentum space, is
inversely proportional to the square of the momentum transfer).

Our final theoretical results for the decay 
rates of the low lying ortho states are
$\tau(1{^3}S_1) = 1.79073(77) \cdot 10^{-12} \, {\rm s}$
and $\tau(2{^3}S_1) = 1.43305(59) \cdot 10^{-11} \, {\rm s}$,
where we have assumed that higher order corrections contribute at the
level of $5\,\%$ compared to next--to--leading order corrections.

In this Letter, we also investigate the decay modes of the para system.
The main decay channel of paradimuonium is given by the annihilation
into two (real) photons. The first--order result for the
decay rate can be extracted from the
imaginary part of the next--to--leading order correction to 
positronium hyperfine splitting (Eq. (\ref{ps5})). The imaginary
part in Eq. (\ref{ps5}) 
is entirely due to the virtual two--photon annihilation of the
para state. The decay rate in leading order can be inferred as
$\Gamma^{(0)}(n{^1}S_0) = \alpha^5 \, m_\mu/(2\,n^3)$.
The next--to--leading order corrections are depicted in Fig.
\ref{diagPM}. They are given by radiative corrections in the 
muon lines, by the wave
function correction due to electronic vacuum polarization and
by two--photon annihilation processes with production of an
additional electron--positron pair. We obtain for the respective
contributions,
\begin{equation} \label{para2} 
\Delta \Gamma(1{^1}S_0) = \frac{\alpha}{\pi} \,
\left[ \frac{\pi^2 - 20}{4} + 1.06 +
\left(\frac{4}{3}\,\ln \frac{2\,m_\mu}{m_e} - \frac{16}{9} \right)
\right] \cdot \Gamma^{(0)}(1{^1}S_0) 
= \frac{\alpha}{\pi} \, 4.79 \cdot \Gamma^{(0)}(1{^1}S_0)\,.
\end{equation}
For the next--to--leading order corrections to the
decay of the $2{^1}S_0$ state, we obtain
$\Delta \Gamma(2{^1}S_0) = 
(\alpha/\pi) \, 4.65 \cdot \Gamma^{(0)}(2{^1}S_0)$.
The theoretical values for the decay rates of para states are
$\tau(1{^1}S_0) = 5.9547(33) \cdot 10^{-13} \, {\rm s}$
and $\tau(2{^1}S_0) = 4.7653(25) \cdot 10^{-12} \, {\rm s}$,
where we have estimated the magnitude of higher order 
QED corrections as $5\,\%$ of the corrections considered above.

The last two diagrams in Fig. \ref{diagPM}, which correspond to the
decay ${\sl paradimuonium} \to ee + \gamma$, contribute at the $1\,\%$ level 
to the decay rate of the para system. The respective correction 
contains the large logarithm of the muon to electron mass ratio
(the $4/3\,\ln (2\,m_\mu/m_e) - 16/9$ term in Eq. (\ref{para2})). 
This contribution has been evaluated in two independent
ways. One of the methods employed is standard S-matrix formalim, using
the computer algebra system {\sc Mathematica} \cite{mathematica} for the
analytical calculations.
For a simplified treatment, we observe that the decay rate of paradimuonium into 
two photons can be written as the imaginary part of the virtual 
two--photon annihilation (last two diagrams in Fig. \ref{diagPs}).
In order to obtain a result for the ${\sl paradimuonium} \to ee + \gamma$ 
decay, we consider vacuum polarization insertions in
the photon lines. The result can then be obtained by 
considering correction to the photon propagator due to
electronic vacuum polarization, and by analyzing the decay
of paradimuonium into two photons, one with vanishing mass, the other
having a mass that corresponds to the sum of the four--momenta 
of the emerging electron-positron pair.
More details of this calculation will be presented elsewhere \cite{pre}.

In conclusion we obtain the value of
$E_{\rm hfs}(1S) = 4.23283(35) \cdot 10^{7}\,{\rm MHz}$ for 
the hyperfine splitting of the $1S$ states in dimuonium. 
This frequency
is lying in the infrared region and can be produced with e.g. diode lasers.
For the lifetime of the ortho
state $1^3S_1$, we obtain 
$\tau(1{^3}S_1) = 1.79073(77) \cdot 10^{-12} \, {\rm s}$.
The para state $1^1S_0$ has a lifetime of
$\tau(1^1S_0) = 5.9547(33) \cdot 10^{-13} \, {\rm s}$.
Some of the corrections to the decay rate are
state dependent and break the $1/n^3$ scaling of
the corrections found in the positronium system. 
Because they can be expressed
as a correction to the wave function at the origin,
caused by the Uehling potential
(e.g. the fifth diagram in Fig. \ref{diagOM}),
these corrections are more sensitive to details
of the vacuum polarization 
potential around the origin than integral
values like the energy shift. 
Due to the small length scale of the muonic system, 
hadronic vacuum polarization corrections also contribute at the level
of one permille to the decay rate of the ortho state.
It should be noted that the enhancement of the orthodimuonium decay
rate by $1.6\,\%$ due to the electronic vacuum polarization in the far
time--like asymptotic region allows for a sensitive test of QED
(the contribution is given by the third term in Eq. (\ref{ortho2})). This
kinematic region cannot be explored in other atomic systems.

The authors would like to acknowledge helpful
discussions with P. Mohr, V. V. Vereshagin, V. A. Shelyuto,
M. Sander and J. Malenfant.
(U. J.) and (G. S.) would like to 
thank Deutsche Forschungsgemeinschaft 
for continued support (contract no. SO333/1-2).
The work of S. K. and V. I. has been 
supported in part by the Russian
Foundation for Basic Research (grant $\#$95-02-03977).
S. K. is grateful for 
the hospitality extended at the Technical 
University of Dresden.

%
%

%
%

%
%

%
%

\end{document}